\documentclass[a4paper,11pt]{article}
\usepackage{graphicx}
\begin{document}
\bibliographystyle{plain}

\begin{flushleft}
{\Large
Sensitivity of Core-Collapse Supernovae to Neutrino Luminosity\\ 
in Cases of Anisotropic Neutrino Radiation
}
\vspace{1cm}
\\
Yuko Motizuki$^{1}$,
Hideki Madokoro$^{1}$,
and
Tetusya Shimizu$^{1}$
\\
1) RIKEN, Hirosawa 2-1, Wako 351-0198 JAPAN\\
\end{flushleft}
\vspace{0.5cm}
\begin{abstract}
We demonstrate the importance of anisotropic neutrino radiation in 
the mechanism of core-collapse supernova explosions.
Through a new parameter study with a fixed radiation field of neutrinos,
we show that global anisotropy of the neutrino radiation is the
most effective mechanism of increasing the explosion energy when
the total neutrino luminosity is given.
We discuss the reason why, and demonstrate how sensitively the
success of a supernova explosion depends on the neutrino luminosity.    
\end{abstract}

\newcommand{\apj}{Astrophys. J.}
\newcommand{\aap}{Astron. Astrophys.}
\newcommand{\ti}{$^{44}$Ti\ }
\newcommand{\msolar}{$M_{\odot}$\,}
\newcommand{\timestento}[1]{\mbox{$\,\times\,10^{#1}$}}
\newcommand{\ttt}[1]{$\times10^{#1}$}
\newcommand{\gsim}{\mbox{%
 \raisebox{0.3ex}{$>$}\raisebox{-0.7ex}{\test*{-0.8em}$\sim$}}\/}
\newcommand{\lsim}{\mbox{\raisebox{0.3ex}{$<$}\raisebox{-0.7ex}{\hspace*{-0.8em}$\sim$}}\/}

\section{Introduction}

For many years since the first work of 
Colgate \& White (1966) \cite{CW66},
numerical simulations of core-collapse supernova explosions have
been exciting topics.
Here a ``successful" explosion must explain the following observed facts:
\begin{itemize}
\item The explosion energy (observed in SN~1987A): 1.5 $\pm \, 0.5 \times 10^{51}$ ergs.
\item Asymmetry observed by polarization measurements in SN~1987A, SN~1993J, and other 
supernovae.
\item Neutron star mass: 1.4 \msolar.
\item Explosive nucleosynthesis: $^{56}$Ni ($\sim$ 0.07 \msolar for SN~1987A).
\end{itemize}  
So far, no ``successful" explosions have been reported in 1-D
and Multi-Dimensional simulations except work by Wilson and Mayle (e.g. \cite{WM88, WM93};
see, a review by Sato of this volume for the status of numerical 
simulation studies).

In the so-called ``delayed explosion" scenario, an important role of convection,
which is simulated with spherical neutrino radiation field, 
has been widely accepted.  
Here we study the supernova explosion energy with paying special 
attention on {\em anisotropic} neutrino emission and {\em locally intense} neutrino heating
thereby produced.
Such a new idea that anisotropic neutrino radiation
may affect the explosion mechanism itself was first proposed by
Shimizu et al. (1994) \cite{S94}, one of the coauthors in the present study, 
and the effect of anisotropic neutrino radiation on the explosion energy
was intensively studied by Shimizu et al. (2001) \cite{S01}.

Figure~1 depicts the dimensionless entropy contour map \cite{MSM03} 
to explain the overall feature of this idea.
First, we assume that a stronger neutrino flux is emitted in the direction of the rotation 
(polar) axis
of a proto-neutron star than that in the equatorial plane.
This anisotropic neutrino emission, for example, 
may be caused by the rotation of the core of a progenitor star
(see \cite{S01} for other possibilities). 
Then high-entropy hot bubble is created along the polar axis 
behind the primary shock.
Convective motion occurs,  
but the convection becomes very global compared with other models
with spherical neutrino radiation.
The final important feature of our model is that 
the shock wave itself is deformed as seen in Figure~1.
It was emphasized in Shimizu et al. \cite{S01} that
a reasonable degree of anisotropy, l$_{z}$/l$_{x}$ = $\sim$1.05 - $\sim$1.2, 
is enough to increase the explosion energy by a large factor.
In the above, l$_{z}$ and l$_{x}$ are, respectively, the local neutrino flux in the polar
axis and that in the equatorial plane.
(see \cite{S01, MSM03} for details of parameterization of neutrino flux distributions). 
Shimizu et al. (2001) concluded that global anisotropic neutrino radiation
is an {\em alternative} mechanism to revive the shock wave and 
lead to a successful explosion other than so far suggested ``convective trigger".

\begin{figure}[tb]
\begin{center}
\includegraphics[scale=0.6]{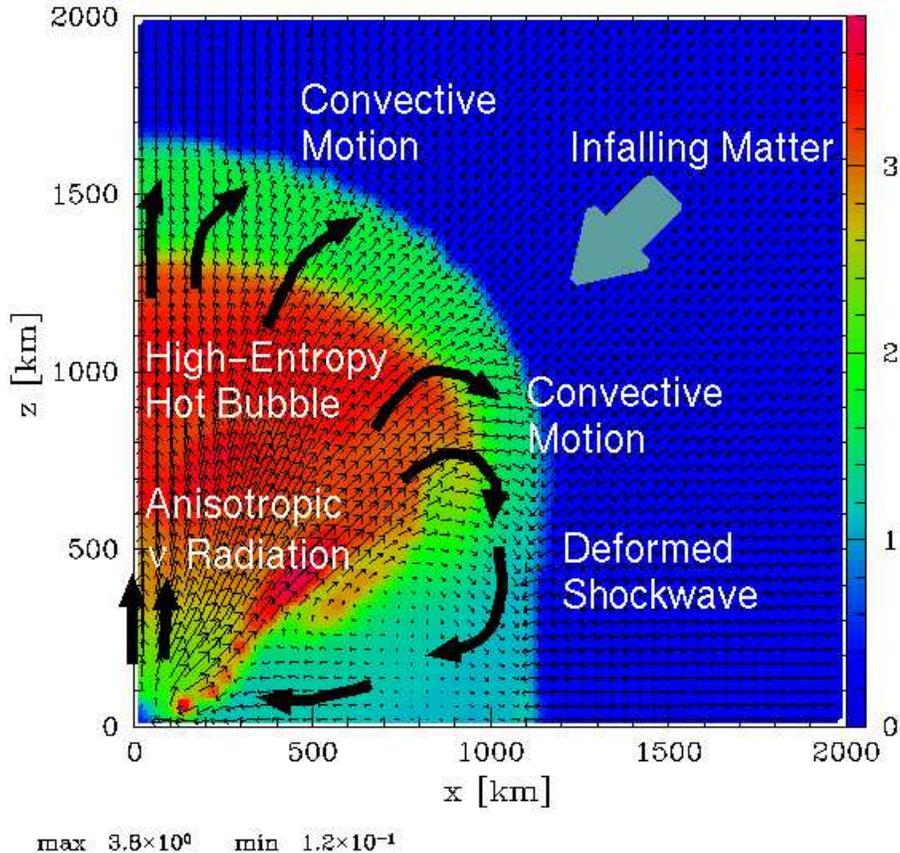}
\end{center}
\caption{
Dimensionless entropy contour map \cite{MSM03} with schematic notations. 
Entropy is calculated for the model of l$_{z}$/l$_{x}$ = 1.1 and 
T$_\nu$ = 4.7 MeV at t = 160 ms after the shock stall;
see section~2 for explanation of this model.
}
\label{fg:figure}
\end{figure}

In this article we are going to show
that global anisotropy of the neutrino radiation is the
most effective mechanism of increasing the explosion energy 
than the models of spherically symmetric and fluctuated neutrino radiation
when the total neutrino luminosity is given.
We discuss the reason why, and demonstrate how sensitively the
success of a supernova explosion depends on the neutrino luminosity. 

Our simulation is performed by solving 2-dimensional hydrodynamic equations
in spherical coordinates. 
A generalized Roe's method is employed to solve the hydrodynamical
equations with general equations of state (EOSs).
Since our method is very rigorous to treat neutrinos above the neutrinosphere,
we put our initial condition (t = 0) when a spherical steady shock is stalled
above the neutrinosphere.  
The details of our numerical technique, together with the EOS and the 
initial condition used, are described in the previous article \cite{S01}.
It is noted here that we have improved the numerical code in the papers 
\cite{S01} and \cite{MSM03} to avoid a numerical error near the pole;
although the error was not serious for the investigation of the explosion energy,
this may affect the results of explosive nucleosynthesis.

\section{Reasoning of Powerful Explosion with Anisotropic Neutrino Radiation}

Figure~2 shows the evolution of the explosion energy, as well as the thermal, 
kinetic, and gravitational energies for models of  l$_{z}$/l$_{x}$ = 1.1 
and T$_\nu$ = 4.7 MeV.
The quantity T$_\nu$ is the neutrino temperature, and we consider here  
only electro-type neutrinos and the total neutrino luminosity 
is fixed (see \cite{MSM03} for details).
In Figure~2, we observe that the difference in the evolution of the explosion energy 
between the globally anisotropic model
and the spherically symmetric model is clearly prominent.

Two dashed lines in the explosion energy (E$_{\rm expl}$) 
in Figure~2 are for the cases when the neutrino field is fluctuated in space
(see \cite{MSM03} for details).
Such neutrino fluctuations may occur due to gravitational oscillation of proto-neutron stars 
as proposed by Burrows et al. \cite{B95}.
The model for the lower dashed line of E$_{\rm expl}$ in Figure~2
has larger mode number of fluctuations in the neutrino flux. 
It is therefore seen that the explosion energy decreases as the mode number of
fluctuations in the neutrino flux increases, and finally approaches that
of spherical explosion.

\begin{figure}[t]
\begin{center}
\includegraphics[scale=0.6]{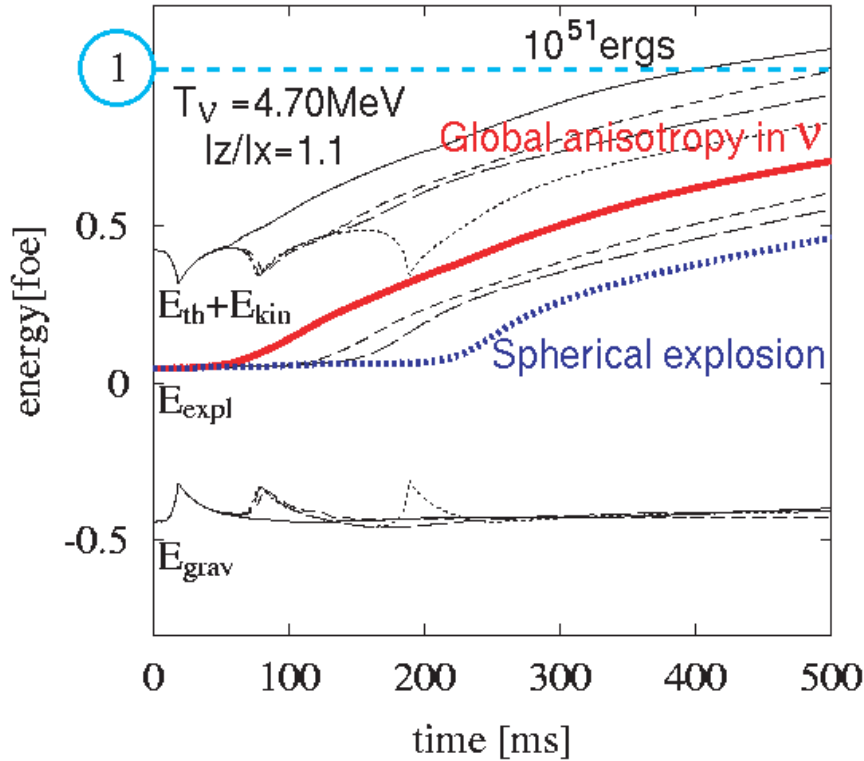}
\end{center}
\caption{
Evolution of thermal and kinetic energy (E$_{\rm th}$+E$_{\rm kin}$), gravitational energy
(E$_{\rm grav}$), and explosion energy (E$_{\rm expl}$) for the model of 
l$_{z}$/l$_{x}$ = 1.1 and T$_\nu$ = 4.7 MeV.  
}
\end{figure}

\begin{figure}[t]
\begin{center}
\includegraphics[scale=0.6]{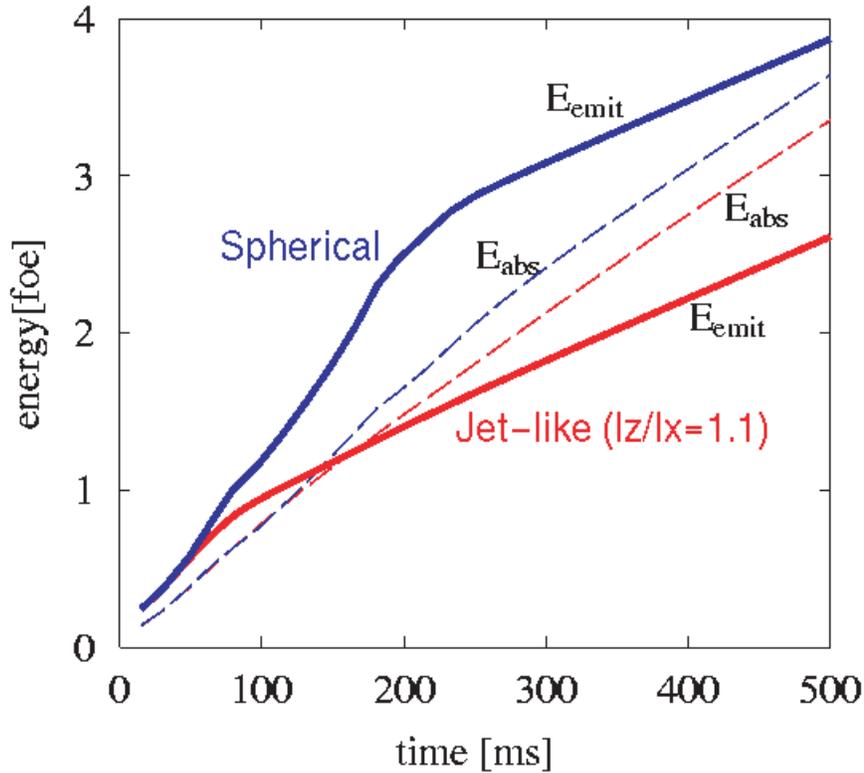}
\end{center}
\caption{
Accumulation of the absorbed and emitted energies due to neutrino 
heating (E$_{\rm abs}$) and cooling (E$_{\rm emit}$), 
respectively, for the anisotropic and spherical models of T$_\nu$ = 4.7 MeV.
Note the difference in the emitted energy.
}
\end{figure}

Let us mention here
the reasoning of this powerful explosion for the globally anisotropic neutrino
radiation model.
Figure~3 shows the accumulation of the absorbed and emitted energies due to 
neutrino heating and cooling for
both globally anisotropic (l$_{z}$/l$_{x}$ = 1.1) and
spherical models.

The balance between heating and cooling determines the 
evolution of the explosion energy.
The crucial point here is the following: 
The neutrino heating is caused by neutrino absorption due to nuclei and
neutrino scattering off electrons and positrons,
and hence the heating rate depends on the {\em neutrino} temperature T$_\nu$: 
It varies as T$_\nu^6$. 
On the other hand, the neutrino cooling is caused by 
neutrino emission due to electron captures and the emission of 
thermal (photo, pair, plasma) neutrinos.  
The cooling rate is then a very sensitive function
of the {\em matter} temperature T, roughly proportional to T$^6$. 

The reasoning of the powerful explosion for the globally anisotropic neutrino
radiation model can be explained as follows.
The locally intense neutrino radiation in our model causes local
heating along the polar axis.
The thermal pressure around the heated matter becomes high, and hence
it pushes part of the shock wave outward.
Namely, the shock revival occurs 
in one direction first along the axis of rotation.
Then the pressure gradient does work along the shock front,
and the shock revival prevails to all directions
(Note the Rankine-Hugoniot condition).
This leads to the earlier shock revival than the spherical model.
Thus the matter temperature behind the shock
decreases rapidly due to expansion.
Accordingly, the neutrino cooling rate rapidly drops.
Heating rate, on the other hand, remains almost unchanged between 
the the anisotropic and the spherical models.
Heating dominates cooling as a result (see Figure~3).
Finally, the explosion energy increases because of the suppression of neutrino
cooling. 

Notice here that the position where the cooling rate suddenly starts to decrease
in Figure~3
(i.e., t~$\sim$~80 ms for the anisotropic, and t~$\sim$~230 ms for the spherical model)
corresponds to the shock revival time, which can be seen in Figure~2 
as the position where the explosion energy suddenly starts to rise up.

\section{Sensitivity of Explosions to Neutrino Luminosities}

\begin{figure}[t]
\begin{center}
\includegraphics[scale=0.6]{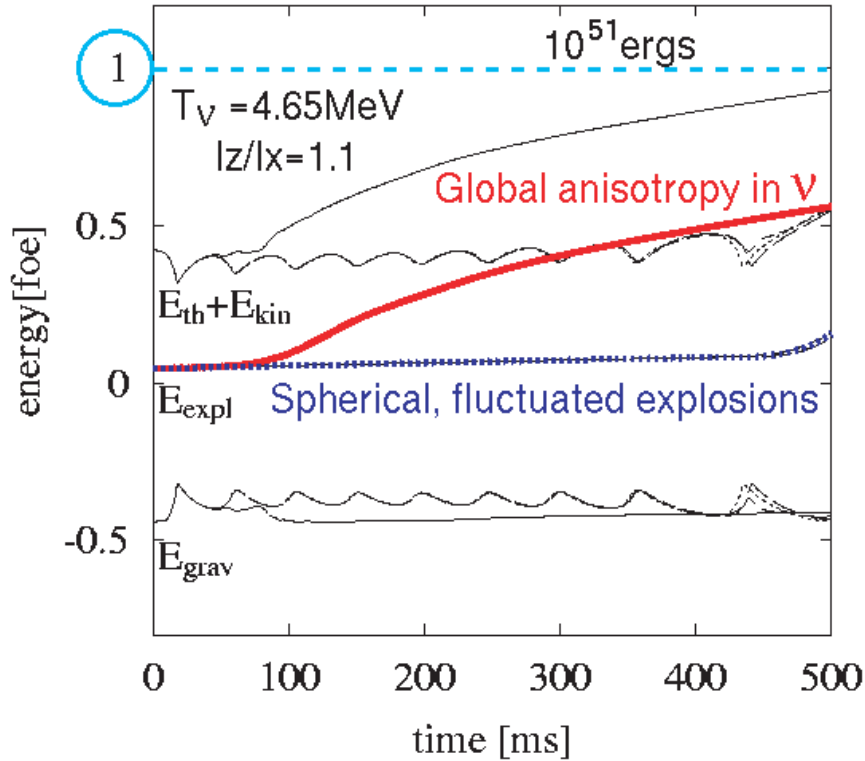}
\end{center}
\caption{
Same as Figure~2 except T$_\nu$ = 4.65 MeV.
}
\end{figure}

In Figure~4 we show a model case with a 
reduced temperature which fails to explode when the neutrino field is 
spherically symmetric and fluctuated, but does produce an adequate explosion when the field 
is reasonably anisotropic (l$_{z}$/l$_{x}$ = 1.1);
the difference between the model with global anisotropy and the other models is extremely
remarkable.

We now see that 
two spherically symmetric models which yield very different explosion outcomes
in Figures 2 and 4, 
but these differ by only 1 \% in neutrino temperature.  
This is a clear demonstration of the sensitivity of the supernova problem 
on the neutrino luminosity and energy.

\section{Concluding Remarks}

It has been discussed that
the globally anisotropic (jet-like) explosion is the most effective mechanism
to increase the explosion energy than spherical and neutrino-fluctuated explosions
when the total neutrino luminosity is given.
While the enhancement of total neutrino luminosity is the most important ingredient
for driving to an energetic explosion as concluded by Janka \& M\"uller 
\cite{JM95, JM96},
the total neutrino luminosity cannot be simply increased to explain the observed
explosion energy.
This is because such treatment often leads to the problem of Ni overproduction,
especially in the case of essentially spherical models. 
We have demonstrated that the 
enhancement of the total neutrino luminosity is not always necessary:
the less luminous the
neutrino radiation is, the more important the neutrino anisotropy is
(compare Figures 2 and 4).
This is a confirmation of the claim by Shimizu et al. \cite{S01}, 
using a different neutrino flux distributions 
and an updated numerical code.

The degree of anisotropy assumed here, l$_{z}$/l$_{x}$ = 1.1, 
can be translated into the rotation period of a proto-neutron star
of $\sim$ 15 ms on the assumption of the Maclaurin spheroid (see \cite{S01}).  
We note that the following saturation properties of the neutrino anisotropy 
has been discussed in the papers \cite{S01, MSM03}; 
the effect of anisotropic neutrino radiation on the explosion energy 
appears with a small value (e.g., l$_{z}$/l$_{x}$ $\sim$ 1.05), 
and the effect saturates
with a certain value (e.g., l$_{z}$/l$_{x}$ $\sim$ 1.2;
the actual value depends on the assumed form of the local neutrino flux).
We emphasize here again that a large anisotropy in the neutrino radiation 
is NOT required for the explosion mechanism.

Finally, we also showed that 
the success of an explosion depends strongly on a subtle difference
in the neutrino energy and luminosities 
 (compare the results of Figures 2 and 4; the temperature difference 
 between 4.70 and 4.65 MeV is only 1 \%). 
What is most important is that anisotropic neutrino radiation,
or local neutrino heating, has been suggested as an alternative key
to the explosion mechanism of supernovae.

\end{document}